\documentclass{appolb}
\usepackage{graphicx}

\begin{document}

\title{Search for novel order in URu$_2$Si$_2$ by neutron scattering
\thanks{Presented at the Strongly Correlated Electron Systems 
Conference, Krak\'ow 2002}
}

\author{M. J. Bull$^1$, B. F{\aa}k$^1$, K. A. McEwen$^2$ and 
J. A. Mydosh$^3$
\address{ $^1$ ISIS, CLRC Rutherford Appleton Laboratory, Chilton, Didcot 
OX11 0QX United Kingdom\\
$^2$ Department of Physics and Astronomy, University College 
London, Gower Street, London WC1E 6BT, United Kingdom\\
$^3$ Kamerlingh Onnes Laboratory, Leiden University, P.O. Box 9504, 2300 
RA Leiden, The Netherlands}}
\date{28 May 2002}
\maketitle

\begin{abstract}
We have made extensive reciprocal space maps in the heavy-fermion 
superconductor URu$_2$Si$_2$ using high-resolution time-of-flight 
single-crystal neutron diffraction to search for signs of a hidden 
order parameter related to the 17.5 K phase transition.  Within the 
present sensitivity of the experiment (0.007 $\mu_B$/U-ion for sharp 
peaks), no additional features such as incommensurate structures or 
short-range order have been found in the $(h0l)$ or $(hhl)$ scattering 
planes.  The only additional low-temperature scattering observed was 
the well-known tiny antiferromagnetic moment of 0.03 $\mu_B$/U-ion.
\end{abstract}

\PACS{75.25.+z;74.70.Tx}   

\section{Introduction}
The nature of the primary order parameter responsible for the entropy 
change at the $T_0=17.5$ K phase transition in the heavy-fermion 
superconductor URu$_2$Si$_2$ continues to be elusive.  The inability 
of the small 0.03 $\mu_B$/U-ion moment arising from long-range 
antiferromagnetic static dipolar order to account for the entropy 
change has led to many theoretical proposals for the nature of a 
primary hidden order parameter, ranging from unconventional 
spin-density waves through to incommensurate orbital 
antiferromagnetism arising from charge currents circulating between 
the uranium ions \cite{chandra}.  To test the validity of some of 
these suggestions, we have used time-of-flight single-crystal neutron 
diffraction to search for additional features that may be present 
below $T_0$.

\section{Experimental}
Time-of-flight single-crystal neutron diffraction is an efficient 
technique that enables data to be simultaneously collected over a wide 
$Q$-range, with each detector scanning a radial trajectory across the 
scattering plane \cite{hagen}.  For these experiments, we have used 
the PRISMA instrument at the UK ISIS neutron spallation source in its 
diffraction mode \cite{prisma} with 30$^\prime$ collimation before the 
detectors, and the detector bank centred at a scattering angle 
$2\theta = 41^\circ$ giving an accessible $Q$-range from 0.75 to 5 
\AA$^{-1}$.  Reciprocal space maps are constructed by rotating the 
sample about the normal to the scattering plane in $8^\circ$ steps 
corresponding to the angular width of the detector array.

The sample used is a large, annealed 0.328 g single crystal with 
dimensions 25x5x3 mm$^3$ with the $a$ direction along the longest 
axis.  EPMA measurements have determined the sample to be single phase 
and of the required composition with no impurity inclusions.  
Resistance measurements confirm the presence of the 17.5 K phase 
transition, and the sample has a residual resistivity of
2.3 $\mu\Omega$m.  The sample was oriented with either the $(h0l)$ or 
$(hhl)$ planes in the scattering plane, and was cooled using a helium 
flow cryostat.  The temperature dependence of the integrated intensity 
of the {\bf Q} $=(100)$ magnetic Bragg peak is well described by 
$I(T)=I(0)[1-(T/T_0)^\alpha]$ with $\alpha = 2.92$, in agreement with
other high-quality samples \cite{fak}.

\section{Results}
Reciprocal space maps are produced from the raw time-of-flight data by 
normalisation to the incident flux and the scattering from a standard 
vanadium sample, and then transforming into the reciprocal lattice 
coordinates of the scattering plane.  Subtracting high-temperature (25 
K) data from low-temperature data (4.5 K) for each scattering plane 
leaves only those features arising from the low-temperature ordered 
phase.  In the $(h0l)$ plane, we observe magnetic peaks at $(100)$ and 
$(102)$, whilst in the $(hhl)$ plane we observe peaks at $(111)$ and 
$(113)$, all arising from the well-known {\bf k} $=(001)$ periodic 
structure associated with the long-range antiferromagnetic static 
dipolar order of the uranium ions.  The subtracted reciprocal space 
map for the $(hhl)$ plane is shown in Fig~\ref{rsmap}.  A cut along 
the $(hh0)$ direction through the $(111)$ magnetic Bragg peak is shown 
in Fig.~\ref{hhcut}.  The clear observation of the tiny ordered moment 
illustrates the excellent signal-to-noise ratio of the PRISMA 
instrument in diffraction mode.

From the measured reciprocal space maps and cuts similar to those 
shown in Fig.~\ref{hhcut}, we conclude that within the covered 
$Q$-range of the experiment and within the present accuracy, no 
additional incommensurate structures or any short-range 
order are observed.  Additionally, an upper bound on the intensity of 
any incommensurate features in the scattering planes investigated can 
be set.  From statistical analysis of the background and peak 
intensities, the detection limit of our experiment is around 1/20 of 
the intensity of the $(100)$ magnetic Bragg peak, i.e. less than 
$\sim0.007\ \mu_B$/U-ion.

\section{Discussion}
In a recent paper inspired by the results of NMR and high-pressure 
neutron diffraction experiments, Chandra {\it et al.} \cite{chandra} 
have suggested that the $T_0$ transition may be dominated by the onset 
of incommensurate orbital antiferromagnetism.  The hidden-order phase 
arises from orbital currents circulating around square uranium 
plaquettes in the $a$-$b$ plane, producing a small net moment 
perpendicular to each plaquet (i.e.\ along the $c$ axis).  The orbital 
currents give rise to small incommensurate Bragg peaks (with a rapidly 
decaying $Q^{-4}$ form factor) principally located around {\bf Q} 
$=(qq1)$ with $q\approx 0.22$.  Furthermore, the intensity of these 
incommensurate peaks in the first Brillouin zone is estimated to be 
$\sim$1/50 of the antiferromagnetic dipolar Bragg peak at e.g. {\bf Q} 
$=(100)$.

In the present experiment we have not detected any of these 
signatures.  In particular, Fig.~\ref{hhcut} shows no signs of any 
peaks at $(hh1)$ with $h=0.22$ or $h=0.78$.  While the experimental 
sensitivity is similar to the predicted intensity, the rapidly 
decreasing form factor could play a role at the relatively large $Q$ 
values we have investigated.  In fact, kinematic constraints mean that 
the $(001)$ position could not be accessed in the present set-up.  
Also, if the orbital moment couples to the neutron spin in the same 
way as the dipolar moment, i.e. only the component of the spin 
perpendicular to the scattering wave vector {\bf Q} is observed, then 
there would be a further decrease of the intensity, in particular for 
{\bf Q}'s close to the $c$ axis.  Further measurements at smaller 
$Q$-values using smaller scattering angles are envisaged.

\begin{figure}
\caption{Subtracted reciprocal space map for the $(hhl)$ scattering 
plane.  Darker points represent intensity above zero after the 
subtraction, with magnetic Bragg peaks located at $(111)$ and $(113)$.  
At nuclear Bragg positions, e.g.~$(110)$, the subtraction is influenced 
by thermal diffuse scattering and detector saturation, but averages to 
zero.  The band of scattering towards the outer edge arises from the 
aluminium tails of the cryostat.}
\label{rsmap}
\end{figure}

\begin{figure}
\caption{Cut through the $(hhl)$  
subtracted map at $Q=(hh1)$ in a direction ${q\parallel(hh0)}$ with width 
$\Delta q_\perp=\pm 0.5c^*$. The line is a Gaussian fit.}
\label{hhcut}
\end{figure}

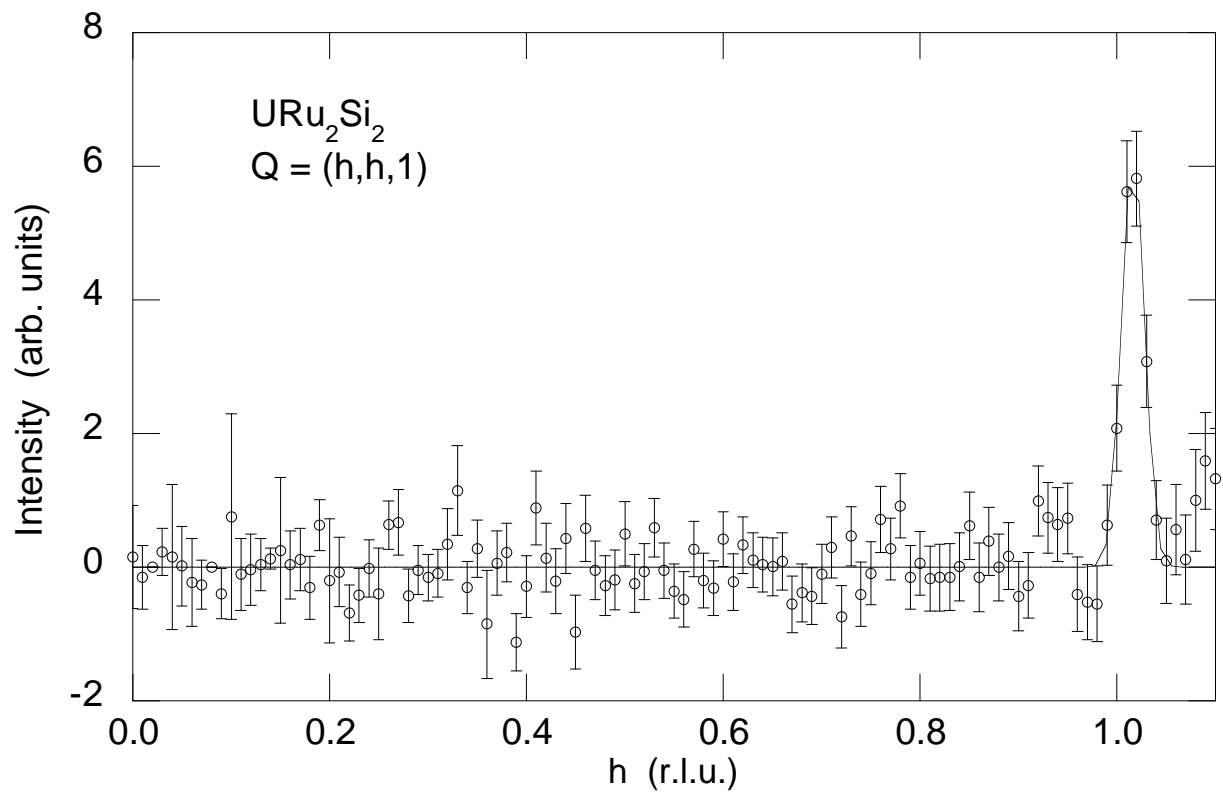
\end{document}